\documentstyle[amssymb,12pt]{article}
\title{{\hfill{\tt Commun.Math.Phys.}}\\{\hfill\tt 81,239-241(1981)}\\
{\hfill}\\Lax Representation for the Systems of S.Kowalevskaya Type}
\author{A.M. Perelomov\\ 
{\small Institute of Theoretical and Experimental Physics, 
117259 Moscow, USSR}}
\date{}
\begin{document}
\maketitle
\begin{abstract}\noindent
We describe a number of dynamical systems that are
generalizations of the S. Kowalevskaya system and admit the Lax
representation.
\end{abstract}

\setcounter{equation}{0}
\noindent
It is well known (see for example [1,2]) that the equations of
motion of a three-dimensional heavy rigid body rotated about a
fixed point is a completely integrable dynamical system only in
the cases of Euler [3], Lagrange [4] and Kowalevskaya [5]. 
In the present note we
describe a number of systems that are of the Kowalevskaya type and
admit the Lax representation\footnote{\,\, Note that the Lax
representation for equations of motion of a rigid body was
discovered first by Manakov [8] for some particular cases.}.

{\bf 1.} Let ${\cal G}$ be the Lie algebra of a group $G$, ${\cal G}^*$ be
the space dual to ${\cal G}$  and $\{x_\alpha \}$ be the coordinates
of a point in the space ${\cal G}^*$. In the space ${\cal F}({\cal G}^*)$ of
smooth functions on ${\cal G}^*$, let us define the Poisson bracket
\footnote{\,\,In this note we use the tensor notations; 
in particular, everywhere
repeated indices imply summation.}
\begin{equation}
\{ f,g\}=C_{\alpha \beta }^\gamma\,x_\gamma \,\partial ^\alpha f
\partial ^\beta g,\qquad \partial ^\alpha =\partial/\partial x_\alpha \,.
\end{equation}
Here $C_{\alpha \beta }^\gamma $ are the structure constants of
the Lie algebra. The space ${\cal F(G^*)}$ is endowed by the above formula
with the structure of the Lie algebra. A dynamical system in ${\cal G}^*$
is determined by a Hamiltonian function $H(x)\in \cal F$, so that the
equations of motion have the form
\begin{equation}
\dot x_\alpha =\{H,x_\alpha \}\,. 
\end{equation}
The coadjoint representation of the group $G$ acts on the space ${\cal G}^*$.
Orbits of this representation  are invariant with respect to an arbitrary
Hamiltonian $H$, and are the phase spaces of the considered systems.

{\bf 2.} Let $G$ be a compact simple Lie group, $K$ be its subgroup such
that the factor space $G/K$ is a symmetric space [6]. Then ${\cal G} =
{\cal K} \oplus {\cal S},\,\,{\cal K}$ is the Lie algebra 
of the group $K$, ${\cal S}$ is the
orthogonal complement to ${\cal K}$ in ${\cal G}$ relative to the
Killing-Cartan form.

It is well known that a certain irreducible representation
$T^0(k)$ of the group $K$ acts on the space ${\cal S}$. Let us
consider an irreducible representation $T(g)$ of the group $G$,
such that under the restriction of this representation on the
subgroup $K$, the irreducible representation $T^0(k)$ is contained
in $T(g)$ with the unit multiplicity. Let $V$ be a vector space in
which the representation $T(g)$ acts. Then it is possible to
define the group $\tilde G=G\cdot V$ that is a semidirect product
of the group $G$ and the abelian vector group $V$. Let us denote
by $V^0$ the subspace  of $V$ on which the representation $T^0(k)$
acts, and by $A$ the projection operator  on the subspace $V^0$.
Let $V^1$ be the orthogonal complement to $V^0\colon V=V^0\oplus
V^1$. Note that the spaces $V^0$ and ${\cal S}$ are isomorphic
because $T(g)$ contains $T^0(k)$ only once. Therefore we have the
decomposition
\[ \tilde{\cal G}={\cal K}\oplus {\cal S}\oplus {\cal V}^0\oplus
{\cal V}^1, \]
and the analogous decomposition for the space $\tilde{\cal G}^*$:
\begin{equation}
\tilde{\cal G}^*={\cal L}\oplus {\cal N}\oplus {\cal P}\oplus {\cal T}.
\end{equation}
Here ${\cal L}={\cal K}^*$, ${\cal N}={\cal S}^*$, ${\cal P}={\cal V}^
{(0)*}$, ${\cal T}={\cal V}^{(1)*}$.
In addition, $\mbox{dim}\,{\cal N}=\mbox{dim}\,{\cal P}=n$, and ${\cal N}$
and ${\cal P}$ are isomorphic relative to the action of the group $K$.

Let $\hat l$ be the matrix of the representation $T$ of the Lie algebra
${\cal G}$, $\hat l=\{ l_{jk}\}$ corresponds to the matrix of the
representation $T^0$ of the algebra ${\cal K}$ acting in the space
${\cal P}$. Let us introduce the matrix, which is important in the sequel,
\begin{equation}
L=A\left[ -\hat l^2+(\gamma \otimes p+p\otimes \gamma )\right] A,
\end{equation}
where $p\in {\cal P}\oplus {\cal T}$ and $\gamma \in {\cal P}$ is the 
constant vector in ${\cal P}$. We identify ${\cal P}$ and ${\cal P}^*$ 
by means of the $K$-invariant scalar product on ${\cal P}$.

The system of S. Kowalevskaya type is defined by the Hamiltonian
\begin{equation}
H=\frac12\,\mbox{tr}\,L=\alpha \,I_2(l)+\beta \,{\cal J}_2(n)+(\gamma p).
\end{equation}
Here $I_2(l)$ and ${\cal J}_2(n)$ are the quadratic functions on
${\cal L}$ and ${\cal N}$ respectively that are invariant relative to the
coadjoint representation of the group $K$ and $p\in {\cal P}$.

{\bf 3.} Let us consider in detail the case: $G=SO(n+1)$, $K=SO(n)$,
$\tilde G=G\cdot V=E(n+1)$ is the motion group of the $(n+1)$-dimensional
Euclidean space $V={\Bbb R}^{n+1}$, $V^0={\Bbb R}^n$, $V^1={\Bbb R}^1$.
Let $\hat l_{jk}=-\,\hat l_{kj}$ and $\hat p_m$ $(j,k=1,\ldots ,(n+1))$
be the standard basis in the space of linear functions
on ${\cal G}^*$ and ${\cal V}^*$ respectively with the standard Poisson
brackets.

The Hamiltonian (5) takes now the form
\begin{equation}
2H=2\sum _{j<k}^nl_{jk}^2+\sum _{j=1}^nn_j^2+2\sum _{j=1}^n
\gamma _jp_j. \end{equation}

\noindent{\bf Theorem.} {\em The equations of motion} (2) {\em of
the system with Hamiltonian} (6) {\em are equivalent to the Lax
equation}
\begin{equation}
\dot L=[L,M], \end{equation}
{\em where} $L$ {\em is given by the formula} (4), {\em and} $M=c\, 
l=c\, A\hat lA$, $c$ {\em is a constant}.

The theorem is verified by a direct calculation.

From the Lax representation (7), it follows immediately that 
the eigenvalues of the matrix $L$ or the quantities 
$I_{2k} =k^{-1}\mbox{tr}\,(L^k)$,
$k=1,\ldots ,n$ are integrals of motion for Eqs. (2). 
Notice $I_2=2H$.

There are also $(n-1)(n-2)/2$ linear integrals related to the
invariance of $H$ with respect to the group $G_0=SO(n-1)$ that
leaves the vector $\gamma $ invariant, and also $([n/2]+1)$
independent polynomials that are invariant with respect to the
coadjoint representation of the group $\tilde G$. Using the theory
of reduction of hamiltonian systems with symmetries [2,7] we may
show that the above system is reduced to a system with
$2n$-dimensional phase space. It can be verifies that integrals
$I_2,I_4,\ldots ,I_{2n}$ are functionally independent  and in
involution. Therefore, the systems under consideration are
completely integrable.

Finally, we wish to note the following facts:
\begin{enumerate}
\item The system considered by S. Kowalevskaya corresponds to $n=2$.
\item Similar results are valid for other symmetric spaces. The simplest ones
are the cases of the spaces of rank one: $SU(n+1)/SU(n)\times U(1)$,
$Sp(n+1)/Sp(n)\times Sp(1)$ and $F_4/SO(9)$.
\item If $p\in {\cal T}$ in formula (5), then we deal with the generalization
of the Lagrange case. In this case, the constants $\alpha $ and $\beta $
may be arbitrary.
\item The results of this section are valid also for similar systems,
related to group $\tilde G=SO(n+2)$.
\end{enumerate}
The detailed presentation of results will be published elsewhere.

\medskip\noindent
{\em Acknowledgement.}\,\,I am grateful to V.Golo for the improvement of 
the language of this note.


\begin{thebibliography}{**}
\bibitem[1]{**} Motion of a rigid body around a fixed point. A collection 
of papers, dedicated to the memory of S. Kowalevskaya. Acad. of Sci. USSR, 
Moscow (1940) (in Russian)
\bibitem[2]{Ar} Arnold, V.I.: Mathematical methods of classical mechanics. 
New York: Springer 1978
\bibitem[3]{Eu} Euler, L.: Decouverte d'un nouveau principe de Mecanique. 
Mem. de l'Acad. des Sciences de Berlin 1758
\bibitem[4]{La} Lagrange, J.L.: Mecanique analytique . Paris 1788
\bibitem[5]{Ko} Kowalevskaya, S.: Acta Math. {\bf 12}, H.2 177--232 (1889)
\bibitem[6]{He} Helgason, S.: Differential geometry and symmetric spaces. 
New York: Acad. Press 1962
\bibitem[7]{Ma} Marsden, J., Weinstein, A.: Rep. Math. Phys. {\bf 5}, 
121--130 (1974)
\bibitem[8]{Ma} Manakov, S.: Funkt. Anal. Priloz. {\bf 10}, 93--94 (1976)

\end{thebibliography}
\end{document}